\documentclass[prd,twocolumn,nopacs,floatfix,amsmath,nofootinbib,amssymb,preprintnumbers]{revtex4-2}
\usepackage{graphicx,subfigure,color,dcolumn,booktabs}
\usepackage{txfonts}
\usepackage{slashed}
\usepackage{epstopdf}
\usepackage{graphicx,color,dcolumn,booktabs}
\usepackage[colorlinks,
            citecolor=blue,
            anchorcolor=red,
            menucolor=red,
            linkcolor=red,
            filecolor=red,
            runcolor=red,
            urlcolor=blue,
            frenchlinks=red]{hyperref}

\graphicspath{{Figures/}}

\begin{document}
\newcommand{\pdlr}{\overset{\leftrightarrow}{\partial}}
\newcommand{\vect}[1]{\boldsymbol{#1}}

\title{Production of charmonium $\chi_{cJ}(2P)$ plus one $\omega$ meson by $e^+e^-$ annihilation}
\author{Ri-Qing Qian$^{1,2,5}$}\email{qianrq21@lzu.edu.cn}
\author{Xiang Liu$^{1,2,3,4,5}$}\email{xiangliu@lzu.edu.cn}
\affiliation{
$^1$School of Physical Science and Technology, Lanzhou University, Lanzhou 730000, China\\
$^2$Lanzhou Center for Theoretical Physics, Key Laboratory of Theoretical Physics of Gansu Province and Frontiers Science Center for Rare Isotopes, Lanzhou University, Lanzhou 730000, China\\
$^3$Key Laboratory of Quantum Theory and Applications of MoE, Lanzhou University, Lanzhou 730000, China\\
$^4$MoE Frontiers Science Center for Rare Isotopes, Lanzhou University, Lanzhou 730000, China\\
$^5$Research Center for Hadron and CSR Physics, Lanzhou University and Institute of Modern Physics of CAS, Lanzhou 730000, China
}

\begin{abstract}
Inspired by the recent observation of $e^+e^-\to \omega X(3872)$ by the BESIII Collaboration, in this work we study the production of the charmonium $\chi_{cJ}(2P)$ by $e^+e^-$ annihilation. We find that the $e^+e^-\to\omega\chi_{c0}(2P)$ and $e^+e^-\to \omega\chi_{c2}(2P)$ have sizable production rates, when taking the cross section data from $e^+e^-\to \omega X(3872)$ as the scaling point and treating the $X(3872)$ as the charmonium $\chi_{c1}(2P)$. Considering that the dominant decay modes of $\chi_{c0}$ and $\chi_{c2}(2P)$ involve $D\bar{D}$ final states, we propose that $e^+e^-\to \omega D\bar{D}$ is an ideal process to identify $\chi_{c0}(2P)$ and $\chi_{c2}(2P)$, which is similar to the situation that happens in the $D\bar{D}$ invariant mass spectrum of the $\gamma\gamma\to D\bar{D}$ and $B^+\to D^+{D}^- K^+$ processes. With the continued accumulation of experimental data, these proposed production processes offer a promising avenue for exploration by the BESIII and Belle II collaborations.
\end{abstract}

\maketitle
\section{Introdution}

Very recently, the BESIII Collaboration announced the observation of the charmoniumlike state $X(3872)$ produced through the $e^+e^-\to \omega X(3872)$ process \cite{BESIII:2022bse}. The line shape of the cross section suggests that the final state $\omega X(3872)$ may involve some nontrivial resonance structures. 
{Indeed, there is already evidence for new resonances in the 4.7-4.8 GeV energy range in several channels. In the $e^+e^-\to K_S^0K_S^0J/\psi$~\cite{BESIII:2022kcv}, a state around $4.71$ GeV is seen with a statistical significance of $4.2 \sigma$. In the measurement of the cross section from $e^+e^-\to D_s^{*+}D_s^{*-}$~\cite{BESIII:2023wsc}, the inclusion of a resonance around 4.79 GeV is necessary to describe the data. A vector charmoniumlike state with a mass of $M=4708^{+17}_{-15}\pm 21$ MeV and a width of $\Gamma=126^{+27}_{-23}\pm30$ MeV is reported in the process $e^+e^-\to K^+K^-J/\psi$~\cite{BESIII:2023wqy} with a significance of over $5\sigma$. The accumulation of data around 4.72 GeV in the $e^+e^-\to \omega X(3872)$ process is consistent with this observed resonance.}
Preliminary analysis indicates that the reported event cluster around 4.75 GeV in the $\omega X(3872)$ invariant mass spectrum aligns well with the predicted $\psi(6D)$ state~\cite{Wang:2020prx}.
However, we should emphasize that the present experimental data cannot conclusively confirm it, and more precise data is required to draw a definite conclusion in near future.

The charmoniumlike state $X(3872)$ has garnered significant attention among the reported charmoniumlike states. Its initial discovery occurred in the $B\to J/\psi\pi^+\pi^- K$ decay channel \cite{Belle:2003nnu}. The low mass puzzle of the $X(3872)$ has sparked extensive discussions, leading to considerations of it as an exotic state (for more in-depth information, refer to review articles \cite{Chen:2016qju,Guo:2017jvc,Olsen:2017bmm,Liu:2019zoy,Brambilla:2019esw,Chen:2022asf}).
In an unquenched picture, the $X(3872)$ can be identified as a $\chi_{c1}(2P)$ state, representing the first radial excitation of the $P$-wave charmonium \cite{Barnes:2003vb,Ortega:2009hj,Kalashnikova:2005ui}. It is part of a triplet of $2P$ states of charmonium, along with the observed charmoniumlike states $X(3915)\equiv \chi_{c0}(2P)$ \cite{Belle:2009and,Liu:2009fe,Duan:2020tsx,ParticleDataGroup:2022pth} and $Z(3930)\equiv \chi_{c2}(2P)$ \cite{Belle:2005rte,Liu:2009fe,Duan:2020tsx,ParticleDataGroup:2022pth}.

The recent observation of the $X(3872)$ in the process $e^+e^-\to \omega X(3872)$ ignites our curiosity and prompts us to explore the production of the remaining $\chi_{c0}(2P)$ and $\chi_{c2}(2P)$ states, accompanied by the $\omega$ meson, through $e^+e^-$ annihilation. This endeavor holds the promise of shedding further light on the properties and characteristics of these fascinating charmoniumlike states.

In this study, we focus on the $e^+e^-\to \omega X(3872)$ process, wherein electrons and positrons annihilate into intermediate vector charmoniumlike state, which subsequently transits into the final state $\omega X(3872)$ through the hadronic loop mechanism. Assuming the intermediate state as a charmoniumlike state, we investigate its decay into the $\omega\chi_{cJ}(2P)$ channels. We calculate the ratios of partial width $\Gamma_{\omega\chi_{cJ}(2P)}$ among different $2P$ states, these ratios then are applied to evaluate the cross sections of the $e^+e^-\to \omega \chi_{c0}(2P)$ and $e^+e^-\to \omega \chi_{c2}(2P)$ processes if taking the experimental cross section data of $e^+e^-\to \omega X(3872)$ as the scaling point with the consideration of  $X(3872)\equiv \chi_{c1}(2P)$. We obtain sizable cross sections of the $e^+e^-\to \omega \chi_{c0}(2P)$ and $e^+e^-\to \omega \chi_{c2}(2P)$ processes comparable to 
the cross section of $e^+e^-\to \omega \chi_{c1}(2P)$, which can provide valuable insights for future experimental searches involved in these two channels.

Contrary to the $\chi_{c1}(2P)$ state, the $\chi_{c0}(2P)$ and $\chi_{c2}(2P)$ states can dominately decay into $D\bar{D}$ final states \cite{Liu:2009fe,Duan:2020tsx}. Hence, we further investigate the processes $e^+e^-\to \omega \chi_{c0}(2P)\to \omega D\bar{D}$ and $e^+e^-\to \omega \chi_{c2}(2P)\to \omega D\bar{D}$. As these two processes share the same final state, the $D\bar{D}$ invariant spectrum can be utilized to identify the $\chi_{c0}(2P)$ and $\chi_{c2}(2P)$ states. However, this task is challenging due to the mass similarity between $\chi_{c0}(2P)$ and $\chi_{c2}(2P)$. Similar to the situation in observing the $\chi_{c0}(2P)$ and $\chi_{c2}(2P)$ in the $B^+\to D^+{D}^- K^+$ decay \cite{LHCb:2020pxc,LHCb:2020bls}, precise data is imperative for successful identification. This work addresses relevant discussions and considerations in this regard. Further experimental advancements and improved data will be essential to discerning these intriguing processes accurately.

This paper is organized as follows. After the introduction, we 
illustrate how to calculate the cross sections of $e^+e^-\to \omega\chi_{cJ}(2P)$ in Sec.~\ref{sec:decay}, where a main task is to introduce the hadronic loop mechanism to study the decay of a vector charmonuimlike $Y$ state decaying into $\omega\chi_{cJ}(2P)$. In Sec.~\ref{sec:DD}, we discuss the contribution of the $\chi_{c0}(2P)$ and $\chi_{c2}(2P)$ states to the $D\bar{D}$ invariant mass spectrum of the $e^+e^-\to\omega\chi_{c0,2}\to \omega D\bar{D}$ channels. Finally, this paper ends with a short summery.

\section{The $e^+e^-\to \omega\chi_{cJ}(2P)$  processes}\label{sec:decay}


The BESIII data of $e^+e^-\to\omega X(3872)$ show that the $\omega\chi_{cJ}$ final state may be attributed to some nontrivial resonance structures, since there exists evidence cluster around the energy range $4.7-4.8$ GeV \cite{BESIII:2022bse}. In fact, similar phenomena have been observed in other processes such as $e^+e^-\to K_S^0K_S^0J/\psi$~\cite{BESIII:2022kcv} and $e^+e^-\to D_s^{*+}D_s^{*-}$~\cite{BESIII:2023wsc}. 
This data accumulation may be due to the predicted charmonium states, specifically the $\psi(7S)$ or $\psi(6D)$~\cite{Wang:2020prx}.
In this context, our primary aim is not to delve into the intricacies of these resonance phenomena. Instead, we briefly highlight that the process $e^+e^-\to\omega X(3872)$ may occur through an intermediate vector charmoniumlike state denoted as $Y$. This serves as a foundation for our subsequent discussions.

\begin{figure}[htbp]
  \centering
  \includegraphics[width=8cm]{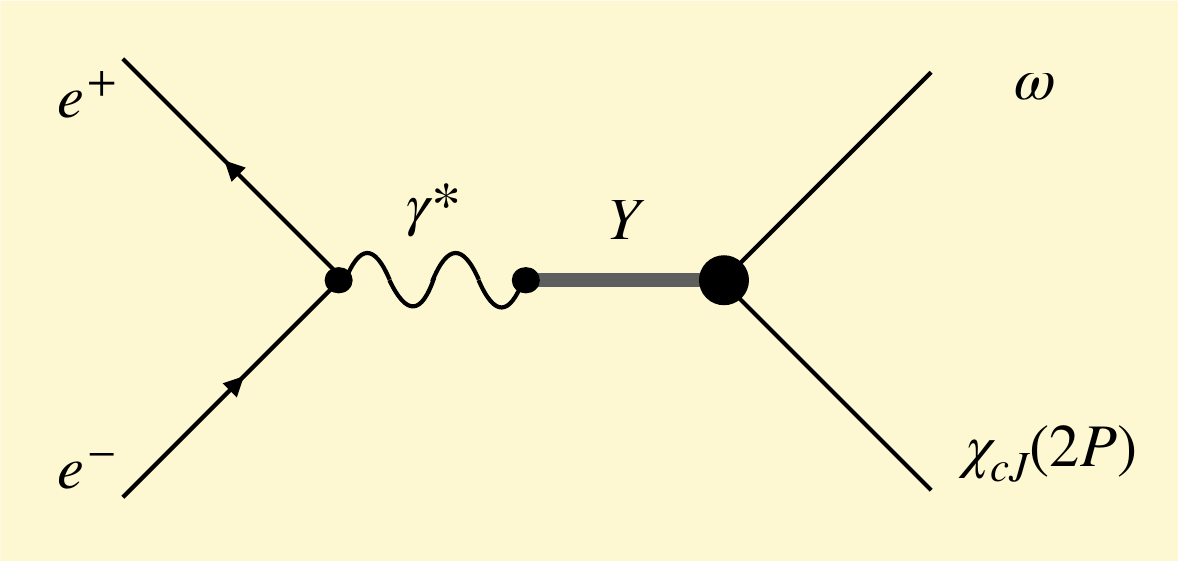}
  \caption{The schematic diagram of the production of the $\chi_{cJ}(2P)$ via $e^+e^-$ annihilation.}\label{fig:tri}
\end{figure}

Upon recognizing that the process $e^+e^-\to\omega\chi_{cJ}(2P)$ proceeds through an intermediate charmoniumlike $Y$ state as depicted in Fig.~\ref{fig:tri}, we can formulate its cross section $\sigma\left(e^+e^-\to Y\to \omega\chi_{cJ}(2P)\right)$ as 
\begin{equation}
\begin{split}
   &\sigma\left(e^+e^-\to Y\to \omega\chi_{cJ}(2P)\right)\\
   &=\frac{12\pi\,\Gamma^{e^+e^-}_Y\Gamma_Y}{|s-m_Y^2+i m_Y\Gamma_Y|^2} \mathcal{BR}\left(Y\to\omega\chi_{cJ}(2P)\right)\,,
\end{split}\label{eq:crosssection}
\end{equation}
which is abbreviated as $\sigma[\omega\chi_{cJ}]$ in the following discussion. Here, $\Gamma^{e^+e^-}_Y$ is the dilepton width of $Y$ and $\mathcal{BR}(Y\to\omega\chi_{cJ}(2P))$ is the decay branching ratios of $Y\to \omega\chi_{cJ}(2P)$. $\Gamma_Y$ and $m_Y$ are resonance parameter of the charmoniumlike $Y$
state.

\begin{figure}[htbp]
  \centering
  \includegraphics[width=8cm]{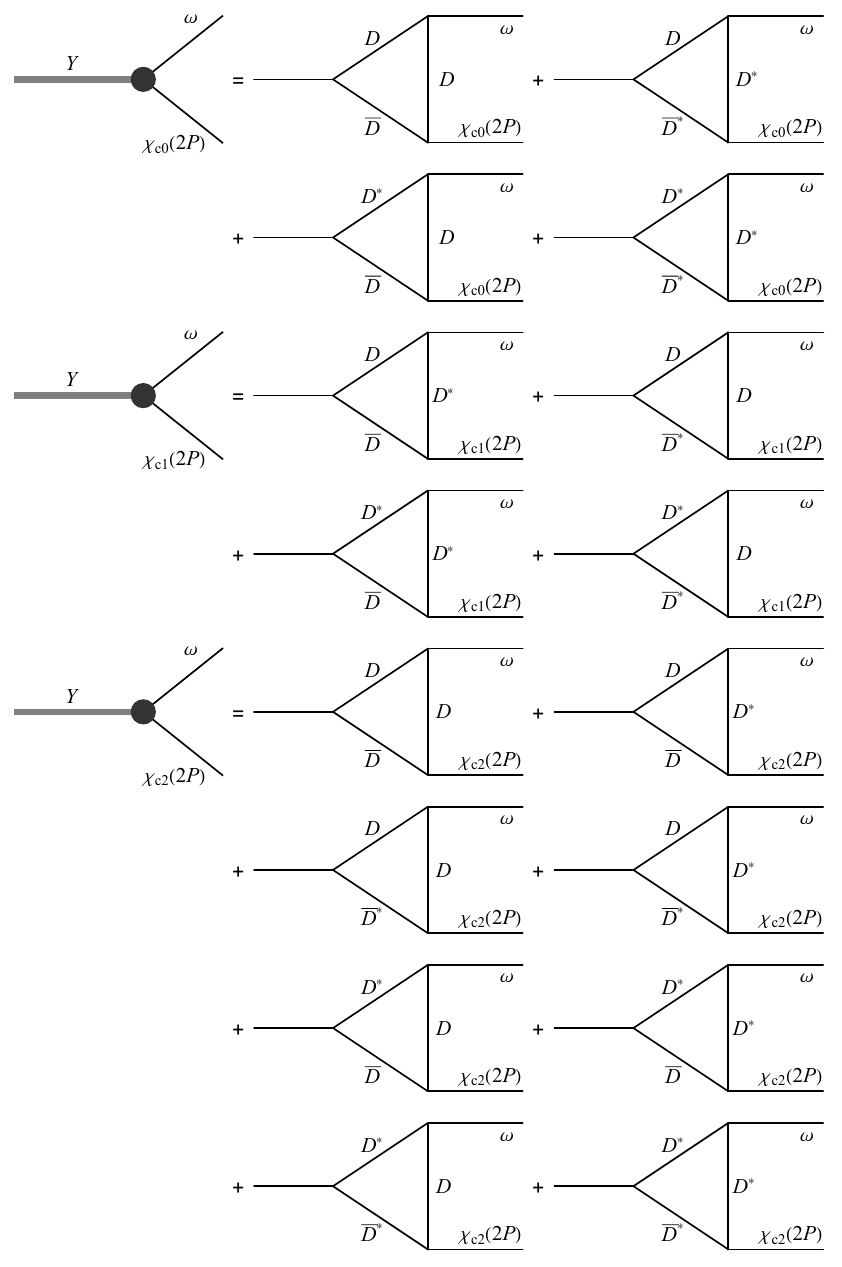}
  \caption{The allowed diagrams of the $Y\to\omega\chi_{cJ}(2P)$ decays by considering the hadronic loop mechanism.}\label{fig:feyn}
\end{figure}

In the subsequent discussion, our focus shifts to the calculation of the decay width of $Y\to\omega\chi_{cJ}(2P)$. To elucidate the peculiar hadronic transition behavior observed in the $\Upsilon$ states, one employed the concept of the hadronic loop mechanism. This effective approach serves as a valuable tool for modeling the coupled channel effect \cite{Liu:2006dq,Liu:2009dr,Zhang:2009kr}, where a loop comprised of bottom mesons acts as a bridge connecting the initial and final states, as illustrated in previous References~\cite{Meng:2007tk,Meng:2008ddd,Meng:2008bq,Chen:2011qx,Chen:2011zv,Chen:2011pv,Chen:2014ccr,Wang:2016qmz,Huang:2017kkg,Huang:2018pmk,Huang:2018cco}. Drawing inspiration from this framework, we can construct schematic diagrams representing the decay process $Y\to \omega\chi_{cJ}(2P)$, as illustrated in Fig.~\ref{fig:feyn}. These decay diagrams  involve charmed meson loops. The general form of the decay amplitudes can be expressed as follows:
\begin{equation}\label{eq:ampgeneral}
  \mathcal{M} = \int \frac{d^4q}{(2\pi)^4} D(q_1,m_{q_1})D(q_2,m_{q_2})D(q,m_q)\mathcal{V}_1\mathcal{V}_2\mathcal{V}_3.
\end{equation}
Here, $D(p,m)=(p^2-m^2)^{-1}$ represents the propagator, and it involves three interaction vertices denoted as $\mathcal{V}_1$, $\mathcal{V}_2$, and $\mathcal{V}_3$ in each diagram. These vertices are associated with their corresponding Lorentz indices, which are appropriately contracted in the calculation.

In order to explicitly formulate the decay amplitudes associated with the diagrams depicted in Fig.~\ref{fig:feyn}, it is essential to define the various interaction vertices, denoted as $\mathcal{V}_i$. These vertices encapsulate the dynamics of the interactions involved. Taking into account  heavy quark symmetry, we can establish the Lagrangian that characterizes the interaction between charmonium states and charmed mesons. These Lagrangians~\cite{Casalbuoni:1996pg,Colangelo:2003sa,Xu:2016kbn} read as
\begin{equation}
\label{eq:Lagrangians}
  \begin{split}
     \mathcal{L}_S & = ig_S \mathrm{Tr}\left[S^{(Q\bar{Q})}\bar{H}^{(\bar{Q}q)}\gamma^\mu \overset{\leftrightarrow}{\partial}_\mu\bar{H}^{(Q\bar{q})}\right]+h.c.\,, \\
     \mathcal{L}_P & = ig_P \mathrm{Tr}\left[P^{Q\bar{Q}\mu}\bar{H}^{(\bar{Q}q)}\gamma^\mu \bar{H}^{(Q\bar{q})}\right]+h.c.\,
  \end{split}
\end{equation}
with $\overset{\leftrightarrow}{\partial}=\overset{\rightarrow}{\partial}-\overset{\leftarrow}{\partial}$, where $\bar{H}$ is related to the doublet field of charmed meson field $H=\left(\frac{1+\slashed{v}}{2}\right)(D^{*\mu}\gamma_\mu+iD\gamma_5)$ by $\bar{H}=\gamma_0H^\dagger\gamma_0$. The $S^{(Q\bar{Q})}$ and $P^{(Q\bar{Q})\mu}$ are the $S$-wave and $P$-wave multiplets of quarkonia,
\begin{equation}\label{eq:swavemul}
  S^{(Q\bar{Q})} = \frac{1+\slashed{v}}{2} \left(\psi^\mu \gamma_\mu-\eta_c \gamma_5\right) \frac{1-\slashed{v}}{2}\,,
\end{equation}

\begin{equation}\label{eq:multiplets}
  \begin{split}
     P^{(Q\bar{Q})\mu} = & \frac{1+\slashed{v}}{2} \left[ \chi_{c2}^{\mu\alpha}\gamma_\alpha
     + \frac{1}{\sqrt{2}}\epsilon^{\mu\alpha\beta\gamma}v_\alpha\gamma_\beta\chi_{c1\gamma}\right.\\
     & \left. + \frac{1}{\sqrt{3}}(\gamma^\mu-v^\mu)\chi_{c0}
     + h_c^\mu\gamma_5 \right] \frac{1-\slashed{v}}{2}\,.
  \end{split}
\end{equation}
The Lagrangian describing the interaction between charmed mesons and light vector mesons is given by~\cite{Casalbuoni:1996pg,Colangelo:2003sa,Li:2021jjt}
\begin{equation}\label{eq:LagV}
  \begin{split}
    \mathcal{L}_V =\, & i\beta\, \mathrm{Tr}\left[ H^{(Q\bar{q})j}v^\mu(-\rho_\mu)^i_j\bar{H}_i^{(Q\bar{q})} \right] \\
     & + i\lambda\, \mathrm{Tr}\left[ H^{(Q\bar{q})j}\sigma^{\mu\nu}F_{\mu\nu}(\rho^i_j)\bar{H}_i^{(Q\bar{q})} \right]\,
  \end{split}
\end{equation}
with $\rho_\mu=ig_V \mathcal{V}_\mu/\sqrt{2}$ and $F_{\mu\nu}=\partial_\mu\rho_\nu-\partial_\nu\rho_\mu+[\rho_\mu,\rho_\nu]$. Here, the light flavor vector meson matrix $\mathcal{V}$ reads as
\begin{equation}\label{eq:vector}
  \mathcal{V} = \left(
                 \begin{array}{ccc}
                   \frac{1}{\sqrt{2}}(\rho^0+\omega) & \rho^+ & K^{*+} \\
                   \rho^- & \frac{1}{\sqrt{2}}(-\rho^0+\omega) & K^{*0} \\
                   K^{*-} & \bar{K}^{*0} & \phi \\
                 \end{array}
               \right)\,.
\end{equation}
We can now get the Feynman rules governing the interaction vertices that are crucial for getting the decay amplitude. These rules have been collected in Table \ref{rule}.

\renewcommand\tabcolsep{0.42cm}
\renewcommand{\arraystretch}{1.50}
\begin{table*}[htbp]
\caption{Feynman rules for the interaction vertexes.}\label{rule}
\begin{tabular}{c|c}\toprule[1pt]\toprule[1pt]
Feynman rules& Feynman rules\\\hline
$\langle D(q_1)\bar{D}(q_2)|\psi(p)\rangle =-g_{\psi DD}\epsilon_\psi^\mu \left( q_{1\mu}-q_{2\mu} \right)$&$\langle D(q_1)\bar{D}^*(q_2)|\psi(p)\rangle=g_{\psi DD^*}\varepsilon_{\alpha\beta\mu\nu}\epsilon_\psi^\mu p^\beta\left(q_2^\alpha-q_1^\alpha\right)\epsilon_{\bar{D}^*}^{*\nu}$
\\
$\langle \bar{D}^*(q_1)D(q_2)|\psi(p)\rangle = g_{\psi DD^*}\varepsilon_{\alpha\beta\mu\nu}\epsilon_\psi^\mu p^\beta\left(q_1^\alpha-q_2^\alpha\right)\epsilon_{\bar{D}^*}^{*\nu}$
&
$\langle D^*(q_1)\bar{D}^*(q_2)|\psi(p)\rangle =  -g_{\psi D^*D^*}\epsilon_\psi^\mu\left(-g_{\alpha\beta}\left(q_{1\mu}-q_{2\mu}\right) +g_{\alpha\mu}q_{1\beta}-g_{\beta\mu}q_{2\alpha}\right)\epsilon_{D^*}^{*\alpha}\epsilon_{\bar{D}^*}^{*\beta} $\\
$ \langle \chi_{c0}(p_2)|D(q)\bar{D}(q_2)\rangle = ig_{\chi_{c0}DD}$&$\langle \chi_{c0}(p_2)|D^*(q)\bar{D}^*(q_2)\rangle = -ig_{\chi_{c0}D^*D^*}\epsilon_{\bar{D}^*}^{\alpha} \epsilon_{D^*}^\beta g_{\alpha\beta}$\\
$\langle \chi_{c1}(p_2)|D^*(q)\bar{D}(q_2)\rangle = -ig_{\chi_{c1}DD^*}\epsilon_{\chi_{c1}}^{*\mu}\epsilon_{D^*}^\alpha g_{\mu\alpha}$&$\langle \chi_{c1}(p_2)|D(q)\bar{D}^*(q_2)\rangle = ig_{\chi_{c1}DD^*}\epsilon_{\chi_{c1}}^{*\mu}\epsilon_{\bar{D}^*}^\alpha g_{\mu\alpha}$\\
$\langle \chi_{c2}(p_2)|D(q)\bar{D}(q_2)\rangle = -ig_{\chi_{c2}DD}\epsilon_{\chi_{c2}}^{*\mu\nu}q_\mu q_{2\nu}$&$\langle \chi_{c2}(p_2)|D^*(q)\bar{D}(q_2)\rangle = ig_{\chi_{c2}DD^*}\varepsilon_{\mu\rho\alpha\beta}\epsilon_{\chi_{c2}}^{*\mu\nu}p_2^\rho q_{\nu}q_2^\beta\epsilon_{\bar{D}^*}^\alpha$\\
$\langle \chi_{c2}(p_2)|D(q)\bar{D}^*(q_2)\rangle = -ig_{\chi_{c2}DD^*}\varepsilon_{\mu\rho\alpha\beta}\epsilon_{\chi_{c2}}^{*\mu\nu}p_2^\rho q_{2\nu}q^\beta\epsilon_{\bar{D}^*}^\alpha$&$\langle \chi_{c2}(p_2)|D^*(q)\bar{D}^*(q_2)\rangle = ig_{\chi_{c2}D^*D^*}\epsilon_{\chi_{c2}}^{*\mu\nu}\epsilon_{\bar{D}^*}^\alpha\epsilon_{\bar{D}^*}^\beta g_{\mu\alpha}g_{\nu\beta}$\\
$\langle \omega(p_1)D(q)|D(q_1)\rangle = -g_{DD\omega}\epsilon_\omega^{*\mu}\left(q_{1\mu}+q_\mu\right)$&$\langle \omega(p_1)D^*(q)|D(q_1)\rangle = -2f_{DD^*\omega}\varepsilon_{\mu\nu\alpha\beta}\epsilon_\omega^{*\mu}p_1^\nu\left(q_1^\beta+q^\beta\right)\epsilon_{D^*}^{*\alpha}$\\
$\langle \omega(p_1)D(q)|D^*(q_1)\rangle = 2f_{DD^*\omega}\varepsilon_{\mu\nu\alpha\beta}\epsilon_\omega^{*\mu}p_1^\nu\left(q^\beta+q_1^\beta\right)\epsilon_{D^*}^\alpha$&$\langle \omega(p_1)D^*(q)|D^*(q_1)\rangle =  \;\epsilon_\omega^{*\mu}\bigg(g_{D^*D^*\omega}g_{\alpha\beta}\left(q_{1\mu}+q_\mu\right) $\\
&$+4f_{D^*D^*\omega}\left(g_{\mu\beta}p_{1\alpha}-g_{\mu\alpha}p_{1\beta}\right)\bigg)\epsilon_{D^*}^\alpha\epsilon_{D^*}^{*\beta}$\\
\bottomrule[1pt]\bottomrule[1pt]
\end{tabular}
\end{table*}

The decay amplitudes of the process $Y\to\omega\chi_{cJ}(2P)$ can be readily derived by employing the interaction vertices specified earlier. To illustrate this, let's explicitly express the amplitudes for the specific case of $Y\to\omega\chi_{c1}(2P)$, namely: 
\begin{equation}
  \begin{split}
     \mathcal{M}^{(a)}_{\omega\chi_{c1}} = & \; i^3\int \frac{d^4q}{(2\pi)^4} \frac{1}{q_1^2-m_D^2}\frac{1}{q_2^2-m_D^2}\frac{-g^{\delta\gamma}+q^\delta q^\gamma/m^2_{D^*}}{q^2-m^2_{D^*}} \\
       & \times \left[-g_{\psi DD}\epsilon_\psi^\mu (q_{1\mu}-q_{2\mu})\right]\\
       & \times \left[-2f_{DD^*\omega}\epsilon_{\nu\kappa\delta\sigma}\epsilon_{\omega}^{*\nu}p_k^\kappa(q_1^\sigma+q^\sigma)\right] \\
       & \times \left[-ig_{\chi_{c1}DD^*}\epsilon_{\chi_{c1}}^{*\theta}g_{\theta\gamma}\right]\mathcal{F}^2(q^2)\,,
  \end{split}
\end{equation}
\begin{equation}
  \begin{split}
     \mathcal{M}^{(b)}_{\omega\chi_{c1}} = &\; i^3\int \frac{d^4q}{(2\pi)^4} \frac{1}{q_1^2-m_D^2}\frac{-g^{\lambda\rho}+q_2^\lambda q_2^\rho/m^2_{D^*}}{q_2^2-m^2_{D^*}}\frac{1}{q_2-m_D^2} \\
       & \times \left[g_{\psi DD^*}\epsilon_\psi^\mu p^\sigma(q_2^\kappa-q_1^\kappa)\right] \left[ g_{DD\omega}\epsilon_\omega^{*\nu}(q_{1\nu}+q_\nu) \right] \\
       & \times \left[ig_{\chi_{c1}DD^*}\epsilon_{\chi_{c1}}^{*\theta}g_{\theta\rho}\right]\mathcal{F}^2(q^2)\,,
  \end{split}
\end{equation}
\begin{equation}
  \begin{split}
     \mathcal{M}^{(c)}_{\omega\chi_{c1}} = & \; i^3\int \frac{d^4q}{(2\pi)^4} \frac{-g^{\alpha\beta}+q_1^\alpha q_1^\beta/m^2_{m_D^*}}{q_1^2-m^2_{D^*}}\frac{1}{q_2^2-m_D^2} \\
     & \times \frac{-g^{\delta\gamma}+q^\delta q^\gamma/m^2_{D^*}}{q^2-m^2_{D^*}}
       \left[g_{\psi DD^*}\epsilon_{\kappa\sigma\mu\alpha}\epsilon_\psi^\mu p^\sigma(q_1^\kappa-q_2^\kappa)\right] \\
       & \times \left[g_{D^*D^*\omega}\epsilon_\omega^{*\nu}g_{\beta\delta}(q_{1\nu}q_\nu)+ 4f_{D^*D^*\omega}\epsilon_\omega^{*\nu}(g_{\nu\delta}p_{1\beta}-g_{\nu\beta}p_{1\delta})\right] \\
       & \times \left[-ig_{\chi_{c1}DD^*}\epsilon_{\chi_{c1}}^{*\theta}g_{\theta\gamma}\right]\mathcal{F}^2(q^2)\,,
  \end{split}
\end{equation}
\begin{equation}
  \begin{split}
     \mathcal{M}^{(d)}_{\omega\chi_{c1}} = & \; i^3\int \frac{d^4q}{(2\pi)^4} \frac{-g^{\alpha\beta}+q_1^\alpha q_1^\beta/m^2_{D^*}}{q_1^2-m_{D^*}^2}\frac{-g^{\lambda\rho}+q_2^\lambda q_2^\rho/m^2_{D^*}}{q_2^2-m^2_{D^*}}\\
     & \times \frac{1}{q^2-m^2_D}  \left[-g_{\psi D^*D^*}\epsilon_\psi^\mu(-g_{\alpha\lambda}(q_{1\mu}-q_{2\mu})\right.\\
       &\left.+g_{\alpha\mu}q_{1\lambda}-g_{\mu\lambda}q_{2\alpha})\right]  \left[2f_{DD^*\omega}\epsilon_{\nu\kappa\beta\sigma}\epsilon_\omega^{*\nu}p_1^\kappa(q^\sigma+q_1^\sigma)\right]\\ &\times \left[ig_{\chi_{c1}DD^*}\epsilon_{\chi_{c1}}^{*\theta}g_{\theta\rho}\right]\mathcal{F}^2(q^2)\,.
  \end{split}
\end{equation}
In the above amplitude expressions, a form factor $\mathcal{F}^2(q^2)$ is introduced to account for the off-shell effect of the exchanged charmed mesons and to circumvent the divergence issue in the loop integral. This form factor takes on a dipole form defined as $
  \mathcal{F}(q^2) = \left(\frac{m_E^2-\Lambda^2}{q^2-\Lambda^2}\right)^2$,
where $m_E$ and $q$ are the mass and four-momentum of the exchanged mesons, respectively. The cut-off parameter have the form
$\Lambda = m_E+\alpha\Lambda_{QCD}$
with $\Lambda_{QCD}=220\;\mathrm{MeV}$ and $\alpha$ being a dimensionless parameter in the model, which is expected to be of order 1 \cite{Cheng:2004ru}.

The total amplitude is
\begin{eqnarray}
\mathcal{M}^{{\rm{Total}}}_{\omega\chi_{c1}}=4\sum_{i=a,b,c,d}\mathcal{M}_{\omega\chi_{c1}}^{(i)}\,,
\end{eqnarray}
where the factor $4$ comes from the sum over isospin doublet and charge conjugation of charmed meson loops. The differential partial decay width of this decay process is
\begin{equation}\label{eq:difpartialwidth}
  \frac{d\Gamma_{\omega\chi_{c1}}}{d\Omega} = \frac{1}{32\pi^2}\frac{|\vect{p}_1|}{m_Y^2}\left|\mathcal{M}^{\mathrm{Total}}_{\omega\chi_{c1}}\right|^2\,,
\end{equation}
where $\vect{p}_1$ represents the center-of-mass momentum of the final state $\omega$ meson, when the initial state $Y$ is unpolarized, we can integrate the angular distribution out and we have 
\begin{equation}\label{eq:partialwidth}
  \Gamma_{\omega\chi_{c1}} = \frac{1}{3}\frac{1}{8\pi}\frac{|\vect{p}_1|}{m_Y^2}\sum_{pol.}\left|\mathcal{M}^{\mathrm{Total}}_{\omega\chi_{c1}}\right|^2\,,
\end{equation}
where the factor of $1/3$ is a result of the averaging over the polarization of the initial vector state.

Using a similar approach, we can investigate the partial widths for the decays $Y\to\omega\chi_{c0}(2P)$ and $Y\to\omega\chi_{c2}(2P)$.

To calculate the partial widths of the decays $Y\to\omega\chi_{cJ}(2P)$, we need to determine various coupling constants appearing in Table~\ref{rule}. 
The coupling of quarkonium multiplet with charmed mesons in Eq.~\eqref{eq:Lagrangians} shares common coupling constants $g_S$ and $g_P$. 
Thus, the coupling constants $g_{\psi D^{(*)}D^{(*)}}$ and $g_{\chi_{cJ}D^{(*)}D^{(*)}}$ are related to $g_S$ and $g_P$ by~\cite{Li:2021jjt}
\begin{equation}\label{eq:couplingratio}
  \begin{split}
     \frac{g_{\psi DD}}{m_D} =\, & \frac{g_{\psi DD^*}m_\psi}{\sqrt{m_Dm_{D^*}}} = \frac{g_{\psi D^*D^*}}{m_{D^*}} = 2g_S\sqrt{m_\psi}\,, \\
     \frac{g_{\chi_{c0}DD}}{\sqrt{3}m_D} = & \frac{\sqrt{3}g_{\chi_{c0}D^*D^*}}{m_{D^*}} = 2\sqrt{m_{\chi_{c0}}}g_P\,, \\
     g_{\chi_{c1}DD^*} =\, & 2\sqrt{2}\sqrt{m_Dm_{D^*}m_{\chi_{c1}}}g_P\,, \\
     g_{\chi_{c2}DD}m_D =\, & g_{\chi_{c2}DD^*}\sqrt{m_Dm_{D^*}}m_{\chi_{c2}} = \frac{g_{\chi_{c2}D^*D^*}}{4m_{D^*}} = \sqrt{m_{\chi_{c2}}}g_P\,.
  \end{split}
\end{equation}
For the charmed meson couplings to vector meson, we have $g_{\mathcal{DDV}}=g_{\mathcal{D^*D^*V}}=\beta g_V/\sqrt{2}$, $f_{\mathcal{D^*DV}}=f_{\mathcal{D^*D^*V}}/m_{D^*}=\lambda g_V/\sqrt{2}$ with $\beta=0.9$, $\lambda=0.56$ GeV$^{-1}$, $g_V=m_\rho/f_\pi$ and $f_\pi=132$ MeV~\cite{Cheng:1992xi,Yan:1992gz,Wise:1992hn,Burdman:1992gh}. In the calculation, the mass of the intermediate charmonium state is taken as $m_Y=4745$ MeV, and the masses of $\chi_{c0}(2P)\equiv X(3915)$, $\chi_{c1}(2P)\equiv X(3872)$ and $\chi_{c2}(2P)\equiv Z(3930)$ are taken from PDG~\cite{ParticleDataGroup:2022pth}, i.e., $m_{\chi_{c0}(2P)}=3921.7$ MeV and $m_{\chi_{c2}(2P)}=3922.5$ MeV.

\begin{figure}
  \centering
  \includegraphics[width=8cm]{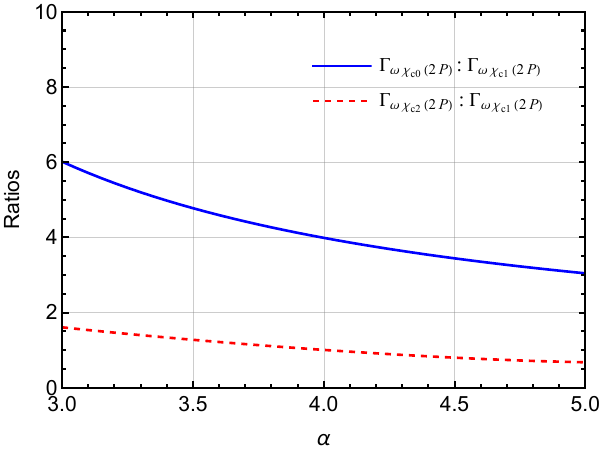}
  \caption{The dependence of the ratios of partial widths of three decay channels $Y\to\omega\chi_{cJ}(2P)$ on $\alpha$. }\label{fig:ratios}
\end{figure}

With these preparations, the ratios of different $\Gamma_{\omega\,\chi_{cJ}}$ can be calculated now. The calculated ratios of different decay channels are shown in Fig.~\ref{fig:ratios}. For our calculations, we consider a range of values for the cut-off parameter $\alpha$ spanning from 3 to 5. It's worth noting that our selection of $\alpha$ falls within a safe range that avoids introducing branch points in the loop integral\footnote{The dipole form factor $\mathcal{F}(q^2)$ resembles a propagator, and the cut-off parameter $\Lambda$ effectively acts as a mass term. This choice prevents the introduction of additional branch cuts in the loop integral, which could occur when specific mass conditions are met, such as $\Lambda+m_{D^{(*)}}=m_{\chi_{cJ}}$.
}.
within this range of $\alpha$, the ratios of the decay widths $\Gamma_{\chi_{cJ}(2P)}$ exhibit only gradual variations in response to changes in the cut-off parameter $\alpha$.
Considering a range of $\alpha$ values from 3 to 5, we obtain the following ratios for the different decay widths:
\begin{equation}\label{eq:ratios}
\begin{split}
   \Gamma_{\omega\,\chi_{c0}(2P)}:\Gamma_{\omega\,\chi_{c1}(2P)} =\, & 3.0\sim 6.0\,, \\
   \Gamma_{\omega\,\chi_{c2}(2P)}:\Gamma_{\omega\,\chi_{c1}(2P)} =\, & 0.7\sim 1.6\,.
\end{split}
\end{equation}
In this context, the partial widths of the three decay channels $Y\to\omega\chi_{cJ}(2P)$ ($J=0,1,2$) exhibit a similar magnitude, with $\Gamma_{\omega\chi_{c0}}$ being slightly greater than $\Gamma_{\omega\chi_{c1}}$ and $\Gamma_{\omega\chi_{c2}}$.

\begin{figure}
  \centering
  \includegraphics[width=8cm]{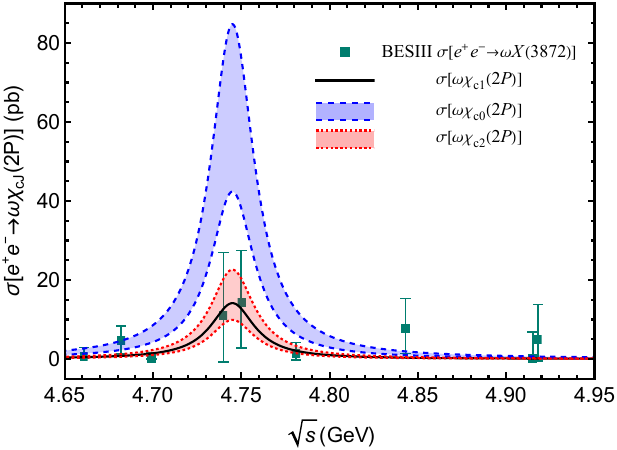}
  \caption{The predicted cross sections of $e^+e^-\to\omega\chi_{c0}(2P)$ and $e^+e^-\to\omega\chi_{c2}(2P)$. Here, we take the experimental data of $e^+e^-\to\omega X(3872)$ \cite{BESIII:2022bse} as the scaling point, treating $X(3872)\equiv\chi_{c1}(2P)$. }\label{fig:sigmachicJ}
\end{figure}

According to Eq.~\eqref{eq:crosssection}, the relative ratios of the production cross sections for different $\chi_{cJ}(2P)$ states can be expressed as
\begin{equation}\label{eq:crosssection1}
  \begin{split}
      & \sigma[\omega\chi_{c0}(2P)]:\sigma[\omega\chi_{c1}(2P)]:\sigma[\omega\chi_{c2}(2P)] \\
      &=\; \Gamma_{\omega\chi_{c0}(2P)}:\Gamma_{\omega\chi_{c1}(2P)}:\Gamma_{\omega\chi_{c2}(2P)}\,.
  \end{split}
\end{equation}
We employ these ratios in Eq.~\eqref{eq:ratios} combined with Eq. (\ref{eq:crosssection1}) to estimate the cross sections for $e^+e^-\to \omega\chi_{c0}(2P)$ and $e^+e^-\to \omega\chi_{c2}(2P)$, utilizing the BESIII data of the cross section of $e^+e^-\to \omega X(3872)$ as the reference scaling point. 
To begin, we assume the existence of an intermediate charmoniumlike state $Y$ with mass $m_Y = 4745 $ MeV and width $\Gamma_Y=30$ MeV. We then fit the experimental data of $e^+e^-\to \omega X(3872)$ using Eq.~\eqref{eq:crosssection} for $e^+e^-\to \omega\chi_{c1}(2P)$. The fitting result is shown in Fig.~\ref{fig:sigmachicJ}. Subsequently, leveraging the obtained ratios of partial widths as described in Eq.~\eqref{eq:ratios}, we proceed to estimate the cross sections for $e^+e^-\to\omega\chi_{c0}(2P)$ and $e^+e^-\to\omega\chi_{c2}(2P)$. These calculations reveal that the cross sections for the $e^+e^-\to\omega\chi_{cJ}(2P)$ processes are of a comparable magnitude, with $e^+e^-\to\omega\chi_{c0}(2P)$ being slightly larger than the others.

Through this study, we underscore the promising prospects for the exploration of the $e^+e^-\to\omega\chi_{c0}(2P)$ and $e^+e^-\to\omega\chi_{c2}(2P)$ processes at BESIII and Belle II in the upcoming years.

\section{The $D\bar{D}$ invariant mass spectrum in $e^+e^-\to\omega\chi_{c0,2}(2P)\to \omega D\bar{D}$}\label{sec:DD}

We should emphasize the importance of analyzing the $D\bar{D}$ invariant mass spectrum to establish the $\chi_{c0}(2P)$ and $\chi_{c2}(2P)$ states. As a good candidate for the charmonium $\chi_{c2}(2P)$ \cite{Liu:2009fe}, charmoniumlike state $Z(3930)$ in $\gamma\gamma\to D\bar{D}$ was reported by the Belle Collaboration~\cite{Belle:2005rte}. Later, Belle observed a charmoniumlike state $X(3915)$ in $\gamma\gamma\to J/\psi\omega$~\cite{Belle:2009and}. The Lanzhou group proposed the $\chi_{c0}(2P)$ assignment to the $X(3915)$ \cite{Liu:2009fe}. In establishing the $\chi_{c0}(2P)$ and $\chi_{c2}(2P)$ states, we should answer several serious questions~\cite{Guo:2012tv,Olsen:2014maa}: 1) Why is the signal of the $X(3915)$ signal missing in the experimental data of the $D\bar{D}$ invariant mass spectrum from $\gamma\gamma\to D\bar{D}$? 2) Why is the mass gap between the $X(3915)$ and $Z(3930)$ so small? Faced with these questions, the Lanzhou group pointed out that the measured $D\bar{D}$ invariant mass spectrum of $\gamma\gamma\to D\bar{D}$ may contain both the $\chi_{c0}(2P)$ and $\chi_{c2}(2P)$ signals \cite{Chen:2012wy}. And the small mass gap between  the $X(3915)$ and $Z(3930)$ can be well explained by considering the coupled channel effect and node effect \cite{Duan:2020tsx}, especially the narrow width of the $\chi_{c0}(2P)$ can also be explained. In 2020, the LHCb Collaboration found the $\chi_{c0}(2P)$ and $\chi_{c2}(2P)$ in the $B^+\to D^+D^-K^+$ decay~\cite{LHCb:2020pxc,LHCb:2020bls}, where both $\chi_{c0}(2P)$ and $\chi_{c2}(2P)$ were observed in the $D^+D^-$ invariant mass spectrum, confirming these predicted properties of the $\chi_{c0}(2P)$ and $\chi_{c2}(2P)$ \cite{Liu:2009fe,Chen:2012wy,Chen:2013yxa,Duan:2021bna}. From this brief review, we can realize that the $D\bar{D}$ invariant mass spectrum plays an important role in deciphering the nature of the $Z(3930)$ and $X(3915)$ and in establishing the $\chi_{c0,2}(2P)$ states.

More observations of the $\chi_{c0}(2P)$ and $\chi_{c2}(2P)$ will help us understand the properties of $\chi_{cJ}(2P)$ better, and $e^+e^-\to\omega\chi_{cJ}(2P)$ seems to be a promising process. The dominant decay channels of $\chi_{c2}(2P)$ are $D\bar{D}$ and $(D\bar{D}^*+h.c.)$ \cite{Liu:2009fe,Chen:2012wy,Chen:2013yxa,Duan:2021bna}, and the $D\bar{D}$ channel can account for $60\%$ of its decay width~\cite{Chen:2013yxa}. For $\chi_{c0}(2P)$, $D\bar{D}$ is the only allowed open charm decay channel \cite{Liu:2009fe,Chen:2012wy,Chen:2013yxa,Duan:2021bna}. So the $e^+e^-\to\omega D\bar{D}$ is a good process to observe $\chi_{c0}(2P)$ and $\chi_{c2}(2P)$ simultaneously. 

Utilizing the calculated ratios of $\Gamma_{\omega\chi_{cJ}(2P)}$ and considering the $D\bar{D}$ decay channel of $\chi_{c0}(2P)$ and $\chi_{c2}(2P)$, we can make a rough prediction of the cross section of $e^+e^-\to\omega D\bar{D}$. As in the previous section, assuming that there is a charmoniumlike state at the position $m_Y=4.75$ GeV, the cross section of $e^+e^-\to\omega D\bar{D}$ at $\sqrt{s}=m_Y$ is estimated to be
\begin{equation}
\begin{split}
  \sigma\left(e^+e^-\to\omega\chi_{c0}(2P)\to\omega D\bar{D}\right) & = (43\sim86)\, \mathrm{pb}\,,\\
  \sigma\left(e^+e^-\to\omega\chi_{c2}(2P)\to\omega D\bar{D}\right) & = (6\sim14)\, \mathrm{pb}\,.
\end{split}
\end{equation}

\begin{figure}[htbp]
  \centering
  \includegraphics[width=8cm]{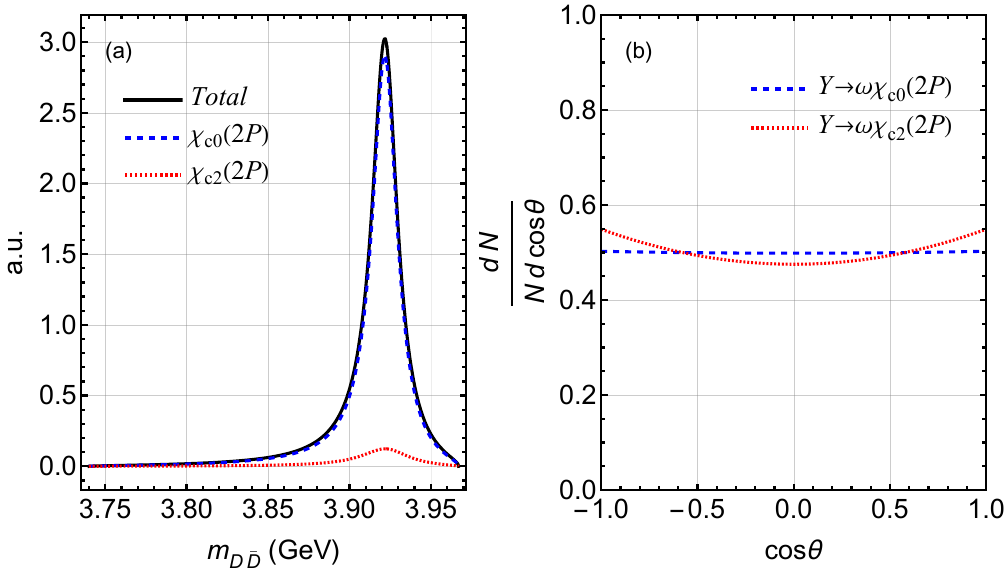}
  \caption{(a) The predicted invariant mass spectrum of $D\bar{D}$ of $e^+e^-\to\omega D\bar{D}$. (b) The calculated angular distributions of $Y\to\omega\chi_{c0,2}(2P)$. Here, the results are properly normalized.}\label{fig:wDD}
\end{figure}

The corresponding $D\bar{D}$ invariant mass spectrum of $e^+e^-\to\omega D\bar{D}$ is shown in Fig.~\ref{fig:wDD}\,(a). Here, the ratios of the partial width of different $\omega\chi_{cJ}(2P)$ channels are taken to be $\Gamma_{\omega\chi_{c0}(2P)}:\Gamma_{\omega\chi_{c1}(2P)}\approx 4.0$ and $\Gamma_{\omega\chi_{c2}(2P)}:\Gamma_{\omega\chi_{c1}}(2P)\approx 1.0$, these values serve as a typical value of our predicted range in Eq.~\eqref{eq:ratios}. The partial width of the $D\bar{D}$ channel of the $\chi_{c0}(2P)$ and $\chi_{c2}(2P)$ channels are set to be $100\%$ and $60\%$, respectively. The small mass gap, $m_{\chi_{c0}(2P)}-m_{\chi_{c2}(2P)}<5$ MeV according to the LHCb measurement~\cite{LHCb:2020pxc,LHCb:2020bls}, make it not an easy task to distinguish the $\chi_{c0}(2P)$ and $\chi_{c2}(2P)$ states directly from the $D\bar{D}$ channel, we suggest that future experiments like BESIII and Belle II focus on this issue with the accumulation of more precise data.

We also consider the angular distributions of the processes $Y\to\omega\chi_{c0}(2P)$ and $Y\to\omega\chi_{c2}(2P)$, which may be a way to discriminate the $\chi_{c0}(2P)$ and $\chi_{c2}(2P)$ states in this process. The angular distribution of $Y\to\omega\chi_{cJ}(2P)$ will be uniform if the produced charmnoniumlike state $Y$ is unpolarized, as indicated by rotational invariance. But the $J^{PC}=1^{--}$ state produced directly from $e^+e^-$ annihilation is polarized, its polarization state is characterized by the spin density matrix $\rho_Y=\frac{1}{2}\sum_{m=\pm 1}|1,m\rangle\langle1,m|$. As a result of this polarization of the vector state $Y$, the decay process $Y\to\omega\chi_{cJ}(2P)$ will have angular distribution of the form:
$\frac{dN}{N d\mathrm{cos}\theta}\propto 1+\alpha_Y \mathrm{cos}^2\theta$,
where $\theta$ denotes the polar angle of $\omega$ in the center of mass frame of the $e^+e^-$ system and $\alpha_Y$ is a coefficient determined by the decay amplitude. Considering the spin density matrix $\rho_Y$, the angular distribution parameter $\alpha_Y$ can be calculated directly from the differential partial width in Eq.~\eqref{eq:difpartialwidth}.

In Fig.~\ref{fig:wDD}\,(b), we show the angular distribution of $Y\to\omega\chi_{c0,2}(2P)$ at $\sqrt{s}=m_Y$. These angular distributions are given when the cut-off parameter $\alpha$ is taken as a typical value of 4. The angular distribution of the $Y\to\omega\chi_{c0}(2P)$ process is almost uniform, while that of the $Y\to\omega\chi_{c0}(2P)$ process is not. However, the difference between the two types of angular distribution is not obvious, which can be distinguished based on large data example in experiments.

\section{summery}\label{sec:summery}

Building upon the recent observation of $e^+e^-\to\omega X(3872)$ process by the BESIII Collaboration \cite{BESIII:2022bse}, we postulate that the process $e^+e^-\to\omega\chi_{cJ}(2P)$ may occur through an intermediate charmoniumlike state denoted as $Y$. To derive the cross sections for these discussed processes, our primary objective is to calculate the decays $Y\to\omega\chi_{cJ}(2P)$, which are facilitated by the hadronic loop mechanism \cite{Meng:2007tk,Meng:2008ddd,Meng:2008bq,Chen:2011qx,Chen:2011zv,Chen:2011pv,Chen:2014ccr,Wang:2016qmz,Huang:2017kkg,Huang:2018pmk,Huang:2018cco}.

Our results reveal that the widths $\Gamma_{\omega\chi_{cJ}(2P)}$ are of comparable magnitude. Treating the $X(3872)$ to be $\chi_{c1}(2P)$, we further estimate the cross sections for $e^+e^-\to\omega\chi_{c0}(2P)$ and $e^+e^-\to\omega\chi_{c2}(2P)$ by employing the experimental data from $e^+e^-\to\omega X(3872)$ as the reference scaling point. This information strongly suggests that $e^+e^-\to\omega\chi_{c0}(2P)$ and $e^+e^-\to\omega\chi_{c2}(2P)$ could indeed be within the reach of experiments like BESIII and Belle II, especially with enhancements of experimental data.

Given that the $D\bar{D}$ final state is the dominant decay channel for both $\chi_{c0}(2P)$ and $\chi_{c2}(2P)$ \cite{Liu:2009fe,Chen:2012wy}, we further delve into the process $e^+e^-\to\omega\chi_{c0,2}(2P)\to \omega D\bar{D}$. By providing the cross section for $e^+e^-\to\omega\chi_{c0,2}(2P)\to \omega D\bar{D}$, along with information about the $D\bar{D}$ invariant mass spectrum and the angular distribution of $\omega\chi_{c0,2}(2P)$ system, we can effectively reflect the contributions of $\chi_{c0}(2P)$ and $\chi_{c2}(2P)$ to the process $e^+e^-\to \omega D\bar{D}$.

Consequently, we propose that the process $e^+e^-\to \omega D\bar{D}$ may serve as a new avenue for accessing the charmonium states $\chi_{c0}(2P)$ and $\chi_{c2}(2P)$, paralleling these two reported channels such as $\gamma\gamma\to D\bar{D}$ \cite{Belle:2005rte} and $B^+\to D^+D^-K^+$ \cite{LHCb:2020pxc,LHCb:2020bls}.

\begin{acknowledgments}
This work is supported by the China National Funds for Distinguished Young Scientists under Grant No. 11825503, National Key Research and Development Program of China under Contract No. 2020YFA0406400, the 111 Project under Grant No. B20063, the fundamental Research Funds for the Central Universities, the project for top-notch innovative talents of Gansu province, and the National Natural Science
Foundation of China under Grants No. 12247101 and No.
12335001.
\end{acknowledgments}

\bibliography{Bibfile}

\end{document}